\documentstyle[aps,prl,floats,epsfig]{revtex}

\newcommand{\bastar}{\begin{eqnarray*}}
\newcommand{\eastar}{\end{eqnarray*}}
\newskip\humongous \humongous=0pt plus 1000pt minus 1000pt

\newif\ifdtup

\relax
\newcommand{\be}{\begin{equation}}
\newcommand{\ee}{\end{equation}}
\newcommand{\bea}{\begin{eqnarray}}
\newcommand{\eea}{\end{eqnarray}}

\newcommand{\tF}{{\tilde F}}

\newcommand{\bare}{{\bar e}}

\newcommand{\pro}{\partial}

\newcommand{\dfrac}{\displaystyle\frac}
\newcommand{\ba}{\begin{array}}
\newcommand{\ea}{\end{array}}

\newcommand{\nn}{\nonumber}

\newcommand{\bmu}{{\bar \mu}}
\newcommand{\bx}{{\bar x}}
\begin{document}
\twocolumn[\hsize\textwidth\columnwidth\hsize\csname@twocolumnfalse%
\endcsname
\title  {A Non-perturbative Estimate of Vacuum Polarization in QED}
\bigskip

\author{Y. M. Cho$^{1, 2}$ and D. G. Pak$^{2
}$}

\address{
$^{1)}$Department of Physics, College of Natural Sciences, Seoul National University,
Seoul 151-742, Korea  \\
$^{2)}$Asia Pacific Center for Theoretical Physics, 207-43 Cheongryangri-dong, Dongdaemun-gu,
               Seoul 130-012 Korea\\
{\scriptsize \bf ymcho@yongmin.snu.ac.kr,
dmipak@mail.apctp.org} \\ \vskip 0.3cm
}
\maketitle

\begin{abstract}

We present a new estimate of the fine structure constant 
and the $\beta$-function of QED at an arbitrary
scale. Using the non-perturbative
but convergent series expression of the one loop effective action of QED
that has been available recently we make a non-perturbative 
estimate of the running coupling and 
the $\beta$-function, and prove the renormalization group
invariance of the effective action. The contrast between our result
and the perturbative result is remarkable.

\vspace{0.3cm}
PACS numbers: 12.20.-m, 13.40.-f, 11.10.Jj, 11.15.Tk
\end{abstract}

\narrowtext
\bigskip
                           ]

It is well-known that the vacuum polarization makes the coupling constant
scale-dependent. This has best been demonstrated in the perturbative
expansion of quantum field theory. On the other hand 
this vacuum polarization effect can also
be studied with the effective action. Recently we have 
obtained a convergent series expression of
the effective action of QED in one loop approximation \cite{cho1}.
{\it The purpose of this Letter is to present a non-perturbative
estimate of the running coupling and
the $\beta$-function of QED from the effective action, 
and to establish the renormalization group invariance of
the effective action}. Remarkably our estimate
provides a significant improvement over the perturbative result.

The effective action of QED has been studied by Euler and
Heisenberg, and by Schwinger long time ago \cite{eul,sch}.
To derive the effective action one may start from the QED Lagrangian
\bea
& {\cal L} = -\dfrac{1}{4} F_{\mu \nu}^2 + \bar \Psi( i {{/ \,}
\llap D} -m) \Psi, \nn\\
& D_\mu= \pro_\mu +ieA_\mu,
\eea
where $m$ is the electron mass. At one loop level one has
\bea
\Delta S &=& i \ln {\rm Det} (i {{/ \,} \llap D}- m),
\eea
so that the effective action
for an arbitrary constant background (after the modified minimal subtraction)
is given by \cite{cho1}
\bea
&&{\cal L}_{eff} =
-\dfrac{a^2-b^2}{2e^2}{\Big (}1-\dfrac{e^2}{12\pi^2}
\ln\dfrac{m^2}{\mu^2}{\Big)} \nn\\
&& - \dfrac{ab}{4\pi^3} \sum_{n=1}^{\infty}\dfrac{1}{n}
{\Big [}\coth(\dfrac{n \pi b}{a}){\Big (} {\rm ci}(\dfrac{n \pi m^2}{a})
\cos(\dfrac{n \pi m^2}{a}) \nn\\
&& +{\rm si}(\dfrac{n \pi m^2}{a}) \sin(\dfrac{n \pi m^2}{a}){\Big )} \nn\\
&&-\dfrac{1}{2} \coth (\dfrac{n \pi a}{b}) {\Big (} \exp(\dfrac{n \pi m^2}{b})
{\rm Ei}(-\dfrac{n \pi m^2}{b})   \nn \\
&&+ \exp(-\dfrac{n \pi m^2}{b}){\rm Ei}(\dfrac{n \pi m^2}{b}
-i\epsilon){\Big )}{\Big ]},
\eea
where $\mu$ is the subtraction parameter and
\bea
a = \dfrac{e}{2} \sqrt {\sqrt {F^4 + (F \tF)^2} + F^2}, \nn \\
b = \dfrac{e}{2} \sqrt {\sqrt {F^4 + (F \tF)^2} - F^2}. \nn
\eea
Notice that in the pure magnetic and the pure electric background it
reduces to
\bea
&{\cal L}_{eff} = - \dfrac{a^2 }{2e^2}
{\Big (}1-\dfrac{e^2}{12\pi^2} \ln\dfrac{m^2}{\mu^2}{\Big)} \nn\\ 
& - \dfrac{a^2}{4\pi^4} \sum_{n=1}^{\infty}\dfrac{1}{n^2} 
{\Big (} {\rm ci}(\dfrac{n \pi m^2}{a}) \cos(\dfrac{n \pi m^2}{a}) \nn\\
& + {\rm si}(\dfrac{n \pi m^2}{a}) \sin (\dfrac{n \pi m^2}{a}) {\Big )},
\eea
and
\bea
&{\cal L}_{eff} = \dfrac{ b^2}{2e^2}
 {\Big (}1-\dfrac{e^2}{12\pi^2} \ln\dfrac{m^2}{\mu^2}{\Big )} \nn\\
&+ \dfrac{b^2}{8 \pi^4}\sum_{n=1}^{\infty}\dfrac{1}{n^2}
{\Big (}  \exp(\dfrac{n \pi m^2}{b}) {\rm Ei}(-\dfrac{n \pi m^2}{b}) \nn\\
&+ \exp(-\dfrac{n \pi m^2}{b} ) {\rm Ei}(\dfrac{n \pi m^2}{b}
-i\epsilon){\Big )}.
\eea

The above effective action is
expressed in terms of the bare coupling and an arbitrary subtraction
parameter. To discuss the physical implications one must
renormalize it first.
To find the renormalized effective action one must discuss the running
coupling and the $\beta$-function.
For this purpose we start from the effective potential in
the pure magnetic background
\bea
&&{V}_{eff} = \dfrac{a^2}{2e^2} (1-\dfrac{e^2}{12\pi^2} \ln\dfrac{m^2}{\mu^2})
+ \dfrac{a^2}{4\pi^4}  f(x),
\eea
where
\bea
&&f(x) =\sum_{n=1}^{\infty}\dfrac{1}{n^2}{\Big(}{\rm ci}(nx) \cos (nx)
+ {\rm si} (nx) \sin (nx) {\Big)}, \nn\\
&&~~~~~~~~~~~~~~~~~~~~~~~x =\dfrac{\pi m^2}{a}.
\eea
With the effective action at hand we can define the running coupling 
$\bare (\bmu)$ by 
\bea
\dfrac{d^2 V_{eff}}{d a^2}{\Big |}_{a = {\bar \mu}^2} = \dfrac{1}{\bare^2}.
\eea
This definition is different from the one used in the perturbative
QED, but certainly is a legitimate definition that one can adopt 
in gauge theories \cite {savv}.
From the definition we obtain
\bea
\dfrac{1}{\bare^2 } &=& \dfrac{1}{e^2}{\Big (} 1 - \dfrac{e^2}{12 \pi^2}
\ln \dfrac{m^2}{\mu^2} {\Big )} + \dfrac{1}{2 \pi^4}
\zeta_1 (\bar x),
\eea
where
\bea
&\zeta_1 (\bx) = {\Big (} f(x) - x \dfrac{df(x)}{dx} + \dfrac{x^2}{2} 
\dfrac{d^2 f(x)}{dx^2}
{\Big )} {\Big |}_{x=\bx}, \nn\\
& \bx = \dfrac{\pi m^2}{\bmu^2}.
\eea

\begin{figure}
\begin{center}
\psfig{figure=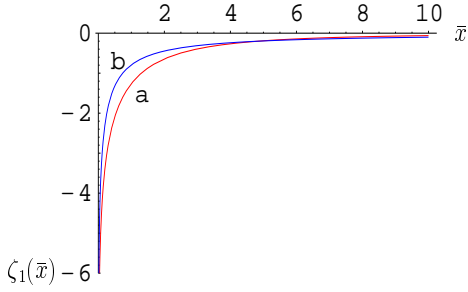, height = 3.5 cm, width = 6.2 cm}
\end{center}
\caption{\label{Fig. 1} The $\bx$-dependence of the vacuum 
polarization function $\zeta_1 (\bar x)$. 
Our result is described by (a) and the perturbative result is 
described by (b).}  
\end{figure}

Notice that with (9) the (running) fine structure constant
is expressed by
\bea
\bar \alpha (\bx)=\dfrac{\alpha}{1 + \dfrac{2 \alpha}
{\pi^3} \zeta_1 (\bx)},
\eea
where $\alpha$ is the asymptotic value of $\bar \alpha$ which we identify
as $\alpha_{\infty} \simeq $ 1/137.036.
This should be compared with well-known vacuum polarization
function of perturbative QED \cite{pesk}
\bea
\zeta_1 (\bx)&&= -\pi^2 \int_{0}^{1}dy~ y(1-y)\ln {\Big (} 1+
\dfrac {\pi y(1-y)}{\bx}{\Big )} \nn\\
&&\simeq \dfrac {\pi^2}{6} ~(~\ln \dfrac{\bx}{\pi}+ 
\dfrac{5}{3}~).~~~~~~~~~~(\bx \ll \pi)
\eea

\begin{figure}
\begin{center}
\psfig{figure=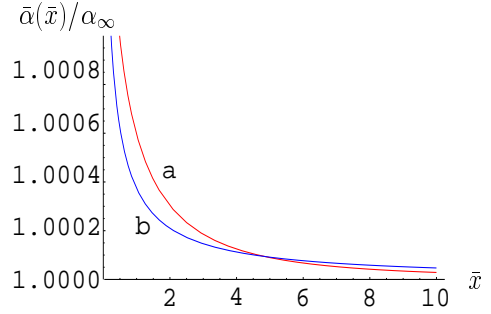, height = 4.25 cm, width = 6.4 cm}
\end{center}
\caption{\label{Fig. 2} The $\bx$-dependence of the fine structure
constant $\bar \alpha (\bar x) / \alpha_{\infty}$. Our result is described by (a),
and the perturbative result is described by (b).}
\end{figure}

The vacuum polarization function $\zeta_1 (\bx)$ and 
the fine structure constant $\bar \alpha (\bx)$
are plotted in Fig. 1 and Fig. 2.
The contrast between our result and that of the perturbative
QED is really remarkable. Obviously our result provides
a significant modification to the perturbative result.
Indeed we find $\bar \alpha (\bmu = M_z) \simeq $ 1/134.555 
from our analysis, which should be
compared with $\bar \alpha (\bmu = M_z) \simeq$ 1/134.647
of the perturbative QED. So we have about 0.07 \% increase at
$M_z \simeq$ 91.189 GeV. Notice that as the energy approaches
to the ultra-violet limit the modification becomes larger.
On the other hand in both cases we still have the Landau
pole at a finite $\bx$, although we can not see this clearly
in the figure. We find that
in our case the position of the Landau pole is given by
$\bx_a \simeq 17.7 ~{\rm x}~ 10^{-562}$, which is of 
the same order as the perturbative
value $\bx_b \simeq 7.4 ~{\rm x}~ 10^{-562}$. So the problem of
the Landau pole in QED does not disappear with our non-perturbative
correction.

From (9) we have the following $\beta$-function,
\bea
\beta(\bx) = \bmu \dfrac{d \bare}{d \bmu}
= \zeta_2 (\bx) ~\bar e^3,
\eea
where
\bea
& \zeta_2 (\bx) = \dfrac{\bx}{2 \pi^4} \dfrac{d \zeta_1 (\bx)}{d \bx}
=\dfrac{\bx^3}{4 \pi^4} \dfrac{d^3 f(\bx)}{d \bx^3} \nn \\
& = \dfrac{ 1}{12 \pi^2} - \dfrac{ \bx^2}{4 \pi^4} 
\sum_{n=1}^{\infty}\Big [1- n \bx \Big ({\rm ci}(n \bx) \sin (n \bx) \nn\\
& - {\rm si} (n \bx) \cos (n \bx) \Big) \Big ].
\eea
On the other hand the perturbative QED gives 
\bea
\zeta_2(\bx)= \dfrac{1}{2\pi}\int_{0}^{1}dy~
\dfrac {y^2(1-y)^2}{\bx+\pi y(1-y)}.
\eea
The function $\zeta_2 (\bx)$ is plotted in Fig. 3.
\begin{figure}
\begin{center}
\psfig{figure=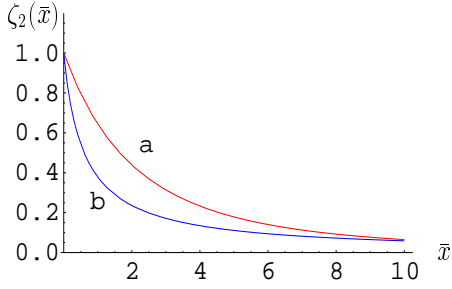, height = 3.8 cm, width = 6 cm}
\end{center}
\caption{\label{Fig. 3} The $\bx$-dependence of $\zeta_2 (\bx)$ which
determines the $\beta$-function. Our result is given in (a) and the
perturbative result is given in (b). Notice that they  are
normalized to one (in the unit $1/12\pi^2$) at the origin.}
\end{figure}
Notice that both our and perturbative $\zeta_2 (\bx)$ 
start from the same familiar value 
$\zeta_2 (0) = 1/12 \pi^2$,
but the discrepancy at a finite $\bx$ is unmistakable.

Using (9) we can express the renormalized effective
potential completely
in terms of $\bar \mu$ and $\bar e$,
\bea
V_{\rm ren} &=& \dfrac{a^2}{2 \bare^2} \Big (1-\dfrac{\bare^2}{2 \pi^4}
     \zeta_1 (\bx) \Big )  + \dfrac{a^2}{4 \pi^4} f(x).
\eea
With this we obtain the Callan-Symanzik equation
which guarantees the renormalization group invariance of the effective
potential
\bea
\Big(\bar\mu\dfrac{\partial}{\partial \bar\mu}
+\beta\dfrac{\partial} {\partial \bar e} \Big)V_{\rm ren}=0.
\eea
Notice that in our notation (3) we have absorbed the coupling constant
$e$ to the gauge field, so that here we have no correction from
the anomalous dimension of the gauge potential in the
Callan-Symanzik equation. 

The Callan-Symanzik equation implies 
that the effective potential (16) is independent of
the scale parameter $\bar x$, so that one should be able to express
the effective potential without the scale parameter. In fact with (11)
we find
\bea
V_{\rm ren} &=& \dfrac{a^2}{2 e_{\infty}^2} 
+ \dfrac{a^2}{4 \pi^4} f(x).
\eea
This tells that one can demonstrate the renormalization 
group invariance of the effective potential directly,
without resorting to the Callan-Symanzik equation.

With this we have the renormalized effective action
which has the manifest renormalization group invariance,
\bea
&&{\cal L}_{\rm ren} =
-\dfrac{a^2-b^2}{2 e_{\infty}^2}
- \dfrac{ab}{4\pi^3}  \sum_{n=1}^{\infty}\dfrac{1}{n}
{\Big [}\coth(\dfrac{n \pi b}{a}) \nn\\
&& {\Big (} {\rm ci}(\dfrac{n \pi m^2}{a})
\cos(\dfrac{n \pi m^2}{a}) 
 +{\rm si}(\dfrac{n \pi m^2}{a}) \sin(\dfrac{n \pi m^2}{a}){\Big )} \nn\\
&&-\dfrac{1}{2} \coth (\dfrac{n \pi a}{b}) {\Big (} \exp(\dfrac{n \pi m^2}{b})
{\rm Ei}(-\dfrac{n \pi m^2}{b})   \nn \\
&&+ \exp(-\dfrac{n \pi m^2}{b}){\rm Ei}(\dfrac{n \pi m^2}{b}
-i\epsilon){\Big )}{\Big ]}.
\eea
Notice that the electric-magnetic duality of the effective action
\cite{cho1} remains intact under the above renormalization.
The real and imaginary parts of the renormalized effective
action are plotted in Fig. 4. In the region shown 
in the figures the quantum fluctuation
provides about 0.1 \% correction to the classical action. 

\begin{figure}
\begin{center}
\psfig{figure=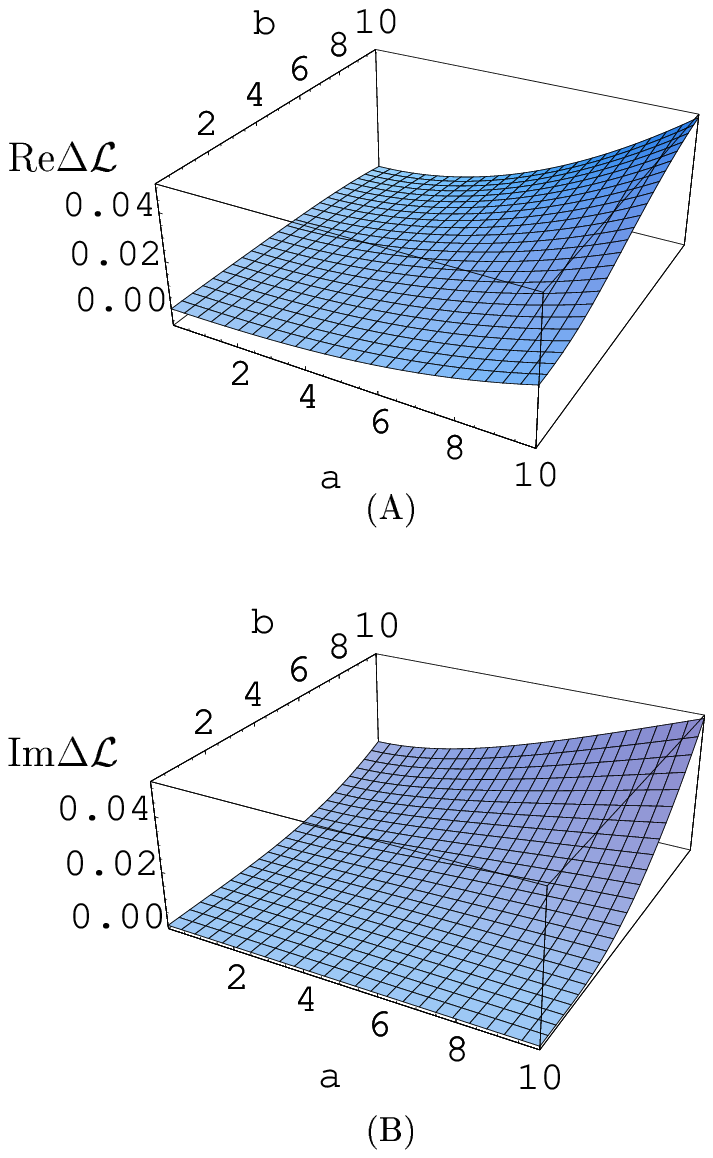, height = 11 cm, width = 6 cm}
\end{center}
\caption{\label{Fig. 4} The one-loop correction to
the dispersive part (A) and the absorptive part (B) of the effective action of QED (in the unit $m^2$). }
\end{figure}

One can obtain the similar results for the scalar QED. In this case
in the pure magnetic and the pure electric background
our one loop effective action reduces to \cite{cho1}
\bea
&{\cal L}_{eff} = - \dfrac{a^2 }{2e^2}
{\Big (}1-\dfrac{e^2}{48\pi^2} \ln\dfrac{m^2}{\mu^2}{\Big)} \nn\\ 
& + \dfrac{a^2}{8\pi^4} \sum_{n=1}^{\infty}\dfrac{(-1)^n}{n^2}  
 {\Big (} {\rm ci}(\dfrac{n \pi m^2}{a}) \cos(\dfrac{n \pi m^2}{a}) \nn\\ 
& + {\rm si}(\dfrac{n \pi m^2}{a}) \sin (\dfrac{n \pi m^2}{a}) {\Big )},
\eea
and
\bea
&{\cal L}_{eff} =  \dfrac{ b^2}{2e^2}
 {\Big (}1-\dfrac{e^2}{48\pi^2} \ln\dfrac{m^2}{\mu^2}{\Big )} \nn\\
& - \dfrac{b^2}{16 \pi^4} \sum_{n=1}^{\infty}\dfrac{(-1)^n}{n^2}
{\Big (}  \exp(\dfrac{n \pi m^2}{b}) {\rm Ei}(-\dfrac{n \pi m^2}{b}) \nn\\
& + \exp(-\dfrac{n \pi m^2}{b} ) {\rm Ei}(\dfrac{n \pi m^2}{b}
-i\epsilon ){\Big )}.
\eea
So the effective potential in the pure magnetic background is given by
\bea
& V_{eff} = \dfrac{a^2 }{2e^2} {\Big (}1-\dfrac{e^2}{48\pi^2}
\ln\dfrac{m^2}{\mu^2}{\Big)} + \dfrac{a^2}{8\pi^4} h(x), \nn\\
& h(x) = \dfrac {}{} \sum_{n=1}^{\infty}\dfrac{(-1)^n}{n^2}
\Big({\rm ci}(nx)\cos(nx)  \nn\\
& + {\rm si} (nx) \sin (nx) \Big). 
\eea
From this with the definition (8) we find
\bea
& \dfrac{1}{\bare^2 } = \dfrac{1}{e^2}{\Big (} 1 - \dfrac{e^2}{48 \pi^2}
\ln \dfrac{m^2}{\mu^2} {\Big )} + \dfrac{1}{4 \pi^4} \eta_1 (\bx), \nn\\ 
& \eta_1 (\bx) = {\Big (} h(x)-x \dfrac{dh(x)}{dx}
+\dfrac{x^2}{2} \dfrac{d^2 h(x)}{dx^2} {\Big)} {\Big |}_{x=\bx}.
\eea
This gives us the following fine structure constant
\bea
\bar \alpha (\bx)=\dfrac{\alpha}{1 + \dfrac{\alpha}
{\pi^3} \eta_1 (\bx)}.
\eea
With this we obtain the $\beta$-function for the scalar QED,
\bea
&\beta(\bar \mu) = \eta_2 (\bx) ~\bar e^3, \nn\\
& \eta_2 (\bx) = \dfrac{\bx}{4 \pi^4} \dfrac{d \eta_1}{d \bx}
 = \dfrac{ 1}{48 \pi^2} + \dfrac{ \bx^2}{8 \pi^4} \sum_{n=1}^{\infty}
  (-1)^n \Big [ 1 - n \bx \nn\\
&\Big ({\rm ci}(n \bx) \sin (n \bx)
- {\rm si} (n \bx) \cos (n \bx) \Big) \Big ].
\eea
From this we finally obtain the following
renormalized effective action
for the scalar QED,
\bea
& {\cal L}_{\rm ren}= -\dfrac{a^2 - b^2}{2 e_{\infty}^2}
+ \dfrac{ab}{8\pi^3} \sum_{n=1}^{\infty}\dfrac{(-1)^n}{n}
{\Big [}{\rm csch} (\dfrac{n \pi b}{a}) \nn\\
& {\Big (} {\rm ci} (\dfrac{n \pi m^2}{a}) \cos (\dfrac{n \pi m^2 }{a})
+ {\rm si}(\dfrac{n \pi m^2 }{a}) \sin(\dfrac{n \pi m^2 }{a}){\Big )} \nn\\
& -\dfrac{1}{2}  {\rm csch}(\dfrac{n \pi a}{b})
{\Big (} \exp (\dfrac{n \pi m^2}{b}) {\rm Ei}(-\dfrac{n \pi m^2}{b})\nn \\
& +\exp (-\dfrac{n \pi m^2}{b}) {\rm Ei } (\dfrac{n \pi m^2}{b}
-i\epsilon){\Big )} {\Big ]},
\eea
which is manifestly invariant under the renormalization
group.

In this Letter we have presented   
a non-perturbative estimate of the vacuum polarization
at an arbitrary scale, and demonstrated the renormalization group
invariance of the one-loop effective action of QED. As far as
we understand it, this is the first time that one has ever
estimated the vacuum polarization non-perturbatively.  
The remarkale contrast between our result and the perturbative result 
is easy to understand theoretically. 
Compared to the perturbative one-loop estimate, our estimate includes
infinitely more Feynman diagrams (one-loop diagrams with 
an arbitrary number of truncated photon legs). 
So, order by order, our estimate is better than the perturbative estimate. 
In this sense our estimate provides 
a definite improvement over the perturbative estimate. 

Certainly one could try to compare our result with experiments.
Here, however, we wish to emphasize the theoretical
importance of our work.  
{\it Our analysis provides a first
non-perturbative estimate of the vacuum polarization in QED which is different
from the existing perturbative estimate. This is really remarkable,
because this is not always the case}. In fact in QCD,  
one can show that the perturbative and non-perturbative estimates
produce identical results, at least at one loop level \cite{cho2}. 

Recently many interesting non-linear phenomena in electrodynamics (e.g.,
the reverse electromagnetic properties of matter, the superluminal
propagation of light, the storage of light, etc.) 
have been studied experimentally \cite{exp}.
Our result should become very useful in 
analyzing these non-linear effects of QED 
\cite{cho3}.

One of the authors (YMC) thanks Professor S. Adler, Professor F. Dyson,
and Professor C. N. Yang for the illuminating discussions.
The work is supported in part by
Korea Research Foundation (KRF-2000-015-BP0072), and by
the BK21 project of Ministry of Education.


\begin{thebibliography}{99}
\bibitem {cho1} Y. M. Cho and D. G. Pak, Phys. Rev. Lett.
{\bf 86}, 1947 (2001);  W. S. Bae, Y. M. Cho, and D. G. Pak, 
Phys. Rev. {\bf D64}, 017303-1 (2001).
\bibitem{eul} W. Heisenberg and H. Euler, Z. Phys. {\bf 98} (1936) 714;
V. Weisskopf, Kgl. Danske Vid. Sel. Mat. Fys. Medd. {\bf 14}, 6 (1936).
\bibitem{sch} J. Schwinger, Phys. Rev. {\bf 82} , 664 (1951).
\bibitem{savv} G. K. Savvidy, Phys. Lett. {\bf B71}, 133 (1977).
\bibitem {cho2} Y. M. Cho and D. G. Pak,
hep-th/0006051, submitted to Phys. Rev. {\bf D}.
\bibitem {exp} D. Smith, W. Padilla, D. Vier, S. Nemat-Nasser, and S. 
Schultz, Phys. Rev. Lett. {\bf 84}, 4184 (2000);
L. Wang, A. Kuzmich, and A. Dogariu, Nature {\bf 406}, 
277 (2000); D. Phillips, A. Fleischhauer, A. Mair, R. Walsworth, 
and M. Lukin, Phys. Rev. Lett. {\bf 86}, 783 (2001).
\bibitem{cho3} Y. M. Cho and D. G. Pak,
to be published.
\end{thebibliography}
\end{document}